\documentclass[12pt, preprint]{aastex}
\usepackage{emulateapj5}  

\def\ergcm2s{~erg cm$^{-2}$ s$^{-1}$ } 
\def\ergs{~erg s$^{-1}$}		
\def\etal{et al.~}		

\def\msun{~M$_{\odot}$}

\def\n4038{~NGC4038/39}		
\def\chandra{{\it Chandra }}

\def\x2{$\chi^{2}$}	



\begin{document}

\title{The Time-Variable Ultra-Luminous X-ray Sources of ``The Antennae'' 
 \\}

\author{ G. Fabbiano$^1$, A. Zezas$^1$, A. R. King$^2$, T. J. Ponman$^3$,
A. Rots$^1$, Fran\c cois Schweizer$^4$}
\affil{$^1$Harvard-Smithsonian Center for Astrophysics, 60 Garden
Street, Cambridge, MA 02138;  gfabbiano@cfa.harvard.edu; azezas@cfa.harvard.edu,
arots@cfa.harvard.edu}
\affil{$^2$Theoretical Astrophysics Group, University of Leicester, Leicester 
LE1 7RH, UK; ark@astro.le.ac.uk}
\affil{$^3$ School of Physics \& Astronomy, University of Birmingham, Birmingham 
B15 2TT, UK; tjp@star.sr.bham.ac.uk}
\affil{$^4$Carnegie Observatories, 813 Santa Barbara St., Pasadena, CA 91101-1292;
 schweizer@ociw.edu}

\shorttitle{\chandra\ The Variable ULXs of ``The Antennae'' Galaxies}
\shortauthors{Fabbiano et al.}
\bigskip

\begin{abstract}
We report the first results of the {\it Chandra} temporal monitoring
of the ultra-luminous X-ray sources (ULXs) in the Antennae galaxies
(NGC~4038/39). Observations 
at four different epochs, covering time scales of 2~years to 2~months,
show variability in seven out of nine ULXs, 
confirming that they are likely to be accreting compact 
X-ray binaries (XRBs). The seven variable ULXs exhibit
a variety of temporal and spectral behaviors: one 
has harder X-ray colors with decreasing luminosity,
similar to the black hole binary Cyg X-1, but 
four other ULXs show the opposite behavior. We suggest
that the latter may be black-hole binaries accreting 
at very high rates.

\end{abstract}

\section{Introduction}

At a distance of 19~Mpc ($H_o = 75$),
NGC~4038/39 have long been studied 
as the nearest examples of a galaxy pair undergoing a major
merger. Recently, high resolution {\it HST} (e.g., Whitmore et al.\ 1999), 
{\it Chandra} (Fabbiano et al.\ 2001),
and {\it VLA} (Neff \& Ulvestad 2000)
observations have produced exquisitely detailed images
of these galaxies, providing an in-depth probe of their young stellar
population. In the X-ray band (0.1--10~keV), the first {\it Chandra}
observation of this system in December 1999 (Fabbiano et al.\ 2001)
revealed a population of  extraordinarily luminous point-like sources.
Nine of these sources, which have hard spectra typical of
X-ray binaries (XRBs), were detected with luminosities
$L_X > 10^{39} \rm ergs~s^{-1}$ $(H_o=75)$, significantly exceeding 
the Eddington luminosity of a spherically accreting neutron star (Zezas et
al.\ 2002a).
This type of Ultra Luminous X-ray source (ULX) has been reported in other
nearby galaxies (e.g., Makishima et al.\ 2000), but never in such copious numbers.
ULXs have been alternatively explained with massive black hole binaries
($M > 10 - 100 M_{\odot}$; e.g. Fabbiano 1989; Makishima et al.\ 2000), young Supernova remnants (e.g.
Fabian \& Terlevich 1996), or
beamed XRBs (King et al.\ 2001). The ULXs of The Antennae appear significantly
displaced from nearby young stellar clusters (Zezas et al.\ 2002b), suggesting that
a significant fraction may not have a very
massive counterpart (Zezas \& Fabbiano 2002).

Given their large number of ULXs and our
thorough knowledge of their stellar populations, The Antennae
provide an exceptional environment for studying ULXs and 
comparing them to the more normal XRB population. 
To this end, a large {\it Chandra} monitoring campaign of The Antennae is under way.
Here we report our first results, by comparing three recent observations
with the original December 1999 data. 

\section{Observations and Analysis}

Table~1 summarizes the log of the four {\it Chandra} ACIS-S (Weisskopf et al.\
2000) observations of the Antennae galaxies discussed in the present Letter,
and lists the net exposure times after background flares
screening. 
%
We corrected the 
astrometry of the December 2001, April 2002, and May 2002 observations, 
following the method of Aldcroft (2002,{\footnote{http://asc.harvard.edu/cal/ASPECT/align\_evt/}}), 
by using bright sources ($>${}$20\sigma$ detections)
within 4\arcmin\ from the optical axis, and referring all these
observations to the one of December 1999, which has
an absolute astrometry good to within $0\farcs5$
(Zezas \etal 2002a). 
From each data set, we then
created images in four different bands, following the
prescriptions by Zezas et al.\ (2002a): Full band (0.3 --
7.0)~keV; soft (0.3 -- 1.0)~keV;
medium (1.0 -- 2.5)~keV; and hard (2.5 -- 7.0)~keV. 
We generated weighted exposure maps for each of the resulting 
images{\footnote{CIAO Thread: http://asc.harvard.edu/ciao/threads/spectral\_weights/}}, 
assuming a 5~keV thermal
bremsstrahlung model with N$_H$=$3.4\times10^{20}~\rm{cm^{2}}$ (corresponding
to the galactic line-of-sight absorbing column; Stark et al.\ 1992),
which is appropriate for the relatively hard emission
of the point-like sources (Zezas et al.\ 2002a, b).
The same model was used to calculate the source fluxes (see below).
We addressed the time-dependent change of the soft 
effective area of ACIS-S 
with the model {\emph{ACISABS}} in
XSPEC{\footnote{http://asc.harvard.edu/cal/Acis/Cal\_prods/qeDeg/}
\footnote{http://www.astro.psu.edu/users/chartas/xcontdir/xcont.html}}.  
We found that, although differences among the last three 
exposures were not significant, their integrated effective areas 
below 1.5~keV were all 
a factor of two smaller than that of the December 1999 exposure.

We used the CIAO  {\emph{wavedetect}} tool for source detection (above
a $3 \sigma$ threshold), 
with the same parameters as 
applied to the Dec 1999 observation (Zezas et al.\ 2002a).
Exposure maps were applied, in order to correct
for spatial variations in the sensitivity of the detector at the 
position of each source between different observations. 
We found that 10 of the sources are more luminous than $10^{39}
\rm ergs~s^{-1}$ in at least one of the observations. These 
sources are the ULXs discovered in the December 1999 data, plus a source
that was present in those data, but at a lower luminosity.
The latter is a super-soft source that is discussed in a separate
Letter (Fabbiano et al., in preparation).
The 9 ULXs are identified in Fig.~1.
We calculated hardness ratios for each source, 
following the procedure and hardness ratio definitions of
Zezas et al.\ 2002a. 
Comparison with the Dec 1999 data (requiring correction for the
effective area degradation) and detailed spectral 
analysis of individual sources will be the subject of future work.


Figure~2 shows the light curves of the nine ULXs (in 10~ks bins), 
labeled with the source numbers from Zezas et al.\ (2002a; see Fig.~1).
Given the error bars, variability in hour or shorter
timescales is not readily detected from the binned data. A KS analysis of
photon arrival times suggests variability within individual observations for
five sources. 
The sources in the top panel appear as luminous ULXs at all 
four epochs, but show a variety of temporal behaviors: Source 11
appears to increase in luminosity during the December 2001 observation 
to reach a peak luminosity of $\sim 5 \times 10^{39} \rm ergs~s^{-1}$ in 
April 2002, and then decline to the level of 
$\sim 1.5 \times 10^{39} \rm ergs~s^{-1}$ in the May 2002 observation, 
a 70\% drop in luminosity; Source 16 exhibits the opposite behavior, 
starting at nearly $ 6 \times 10^{39} \rm ergs~s^{-1}$ in Dec 1999, 
dipping to 40\% of this luminosity in April 2002, and then recovering
in the last observation.  In contrast, Source 29, the nucleus of NGC~4039,
is  steady throughout.  This latter source is
embedded in a luminous extended emission region and has a 
 soft spectrum, which could be due to complex emission from the
nuclear starburst activity (Zezas et al.\ 2002b). The middle panel of Fig.~2
shows three
slowly varying light curves: Sources 37 and 44 show a slight increase
in luminosity from December 1999 to December 2001, and then decline steadily,
while Source 42 shows a continuously declining luminosity.
The maximum luminosity variation for these three sources is $\sim 50\%$.
The sources in the lower panel of Fig.~2 are all fainter:
Source 31 has a
steady luminosity of order $10^{39} \rm ergs~s^{-1}$; Sources 32 and 49
disappear below the detection threshold of $\sim 5 \times 10^{37}
 \rm ergs~s^{-1}$ in April 2002, and reappear at the $10^{38} \rm ergs~s^{-1}$
level in May 2002.  

Figures~3a, b show the tracks of the six more luminous sources in two
X-ray color--color diagrams.
One of the five luminous variable sources (37) has colors suggesting a hardening
of the spectrum with decreasing $L_X$, while all others show the opposite
behavior: 42 and 44 become softer, following their
smoothly decreasing $L_X$; 11 and 16, who undergo a luminosity flare and a dip
respectively, have respectively harder and softer colors corresponding
to these events, although the color changes in 16 are marginal.
Source 29 (the nucleus of NGC~4039) is the only steady source, and has consistent HR3 (S-H/S+H) colors,
with a possible variation in the relative medium band emission in the April
2002 data. The other steady source (31) has consistent colors (not
plotted), but significantly larger error bars.
Sources 32 and 49 are too faint for meaningful color determinations;
their spectra are generally consistent with those of the other
variable ULXs in the December 1999 data (Zezas et al.\ 2002a).

\section{Discussion}

The above results show that all but two of the nine ULXs we detect in the
{\it Chandra} monitoring observations of the Antennae galaxies are variable. 
Of the two non-variable ULXs, one is rather
faint, and the other is the nucleus of NGC~4039. Hence, our results are consistent 
with widespread variability of all non-nuclear ULXs.
Time variability (and when available
spectral variability) suggests that XRBs are the likely counterparts, strengthening
the earlier conclusion  of Zezas et al.\ (2002a, b), which was based mainly on the hard X-ray
spectra of the ULXs.

In only one case do we find the `canonical' high/soft--low/hard
behavior seen in Galactic  black hole binaries such as Cyg X-1.
This behavior has been explained with the dominance of the accretion disk emission
in high state, over that of the innermost spherical
hot flow that may be responsible for the power-law in the thin-disk model
(Shapiro et al.\ 1976), in response to increased accretion rates (e.g., Esin
et al.\ 1998;
Janiuk et al.\ 2000). High/soft--low/hard transitions have also been observed in some ULXs in nearby
galaxies, including M81~X-9 (La~Parola et al.\ 2001), M33~X-8 at the nucleus of M33
(La~Parola et al.\ 2002), and two ULXs in IC~342 (Kubota et al.\ 2001).

In four other ULXs the spectrum
softens with decreasing flux. A recent RXTE monitoring study
has reported this kind of behavior in a few galactic XRBs
(1E~1740.7-2942, GRS~1758-258, GX~339-4, Smith et al.\ 2002; see also the
XMM-Newton results on GRS~1758-258, Miller et al.\ 2002).
Smith et al.\ invoke a two-flow thin disk plus hot halo model.
They suggest that a delayed response of the disk to a drop in
accretion, to which the hot halo would respond immediately,
could cause this low-soft effect in sources with large accretion
disks. They argue that large disks may result from Roche-Lobe overflow 
from a low-mass K star companion.
In their picture, the standard Cyg~X-1 high/soft behavior could be related to the smaller
accretion disk resulting from accretion from a massive companion, that
would respond faster to changes in accretion rate.

Our results are at odds with this scenario. Although ULXs have also been detected in
older stellar systems (E and S0 galaxies, e.g. Angelini et al.\ 2001; Colbert \& Ptak 2002)
the similar flat X-ray luminosity functions of The Antennae
and other star-forming galaxies (Zezas \& Fabbiano 2002; Kilgard et al.\ 2002)
connects the ULXs of The Antennae with the younger stellar population,
suggesting more massive companions than K stars,  in contradiction with the
Smith et al.\ (2002) scenario.
Instead, binaries with very high accretion rates near the black hole may 
emit anisotropically and explain ULXs. This can occur in two situations: 
during the thermal-timescale mass transfer characterizing the later 
stages of a massive XRB (SS433 may be an example) and during 
outbursts of soft X-ray transients (as seen in microquasars).
 While the latter may explain the ULXs in E and S0
galaxies, the former is likely to dominate in The Antennae (King et al.\ 2001;
King 2002).
The X-ray emission in such very high accretion-rate sources
would consist of direct blackbody emission from the
immediate vicinity of the accretor with a temperature of a few keV, and emission
from the central X-rays absorbed and re-emitted by the disk with
slightly lower temperatures (see e.g. Frank et al.\ 2002, eqs 5.44, 5.96). Thus, if 
the accretion rate drops slightly, the spectrum will soften, as it is essentially black body,
resulting in the low/soft behavior. A similar conclusion was reached by Miller
et al.\ (2002),
in their discussion of the low/soft behavior of the galactic microquasar GRS~1758-258.
It may be a relevant point that the average co-added spectrum of the Antennae ULXs
resembles that of galactic microquasars (Zezas et al.\ 2002b). 
In contrast, for much lower accretion rates
the central blackbody source will be too soft to
contribute strongly to the X-ray spectrum, which will be dominated
by the corona, and sources will follow the Cyg~X-1-like low/hard behavior.

In the King et al.\ (2001) model, beamed radiation from a thick
accretion disk, resulting from  thermal time scale mass transfer,
could be responsible for the ULXs. 
The only restriction on the companions in this model is that
they should have radiative envelopes, i.e., be earlier than about
A0. The suggestion that the ULXs of The Antennae are runaway binaries
originating from the young star clusters (Zezas et al.\ 2002b) may
further restrict the companion mass to later than B2, since more
massive systems would have evolved off the main sequence before
reaching the observed average offset from the parent cluster (Zezas \&
Fabbiano 2002).  While a recent paper by Misra \& Sriram (2002) argues
that axisymmetric funnel beaming cannot enhance the apparent luminosity to ULX
values, self-consistent disk modeling shows that the disk must warp,
resulting in a more efficient confinement of the radiation field (see
Fig.~7 of Pringle 1997). As observational support of this, the
galactic microquasar GRS1915+105 reaches ULX luminosities and has an
apparent luminosity well above its Eddington limit. Greiner et al.\
(2001) estimate a distance of 12~kpc. The observed X-ray flux implies an
apparent (isotropic) luminosity of $\sim 7\times 10^{39}$~\ergs, about
3.8 times the Eddington limit corresponding to Greiner et al's
measured dynamical mass of 14 \msun.

\section{Conclusions}
We have found widespread variability of all non-nuclear ULXs in the Antennae 
galaxies. While one of these sources follows a Cyg~X-1
like high/soft--low/hard behavior, four other sources instead become softer
with decreasing luminosity. We suggest that black hole binaries with 
very high accretion rates could explain the low/soft correlation. In all
cases beaming due to thick warped disks could boost the luminosity to
super-Eddington values. 

\acknowledgments

 We thank the CXC DS and SDS teams for their efforts in reducing the data and 
developing the software used for the reduction (SDP) and analysis
(CIAO), and Jon Miller for interesting conversations. 
This work was supported by NASA contract NAS~8--39073 (CXC)  and NASA
Grant G02-3135X.
ARK gratefully acknowledges a Royal Society Wolfson Research Merit Award.

{}

\makeatletter
\def\jnl@aj{AJ}
\ifx\revtex@jnl\jnl@aj\let\tablebreak=\nl\fi
\makeatother
\begin{deluxetable}{ccc}
\tabletypesize{\scriptsize}
\tablecolumns{3}
\tablewidth{0pt}
\tablecaption{Observation log}
\tablehead{ \colhead{OBSID} & \colhead{Date} & \colhead{Net Exposure (sec)}}
\startdata
315 & 1999-12-01 & 75,533 \\
3040 & 2001-12-29 & 63,765  \\
3043 & 2002-04-18 &  60,816 \\
3042 & 2002-05-31 & 67,277 \\
\enddata
\end{deluxetable}

\clearpage

\setcounter{figure}{0}

\begin{figure}
\includegraphics[width=12.0cm]{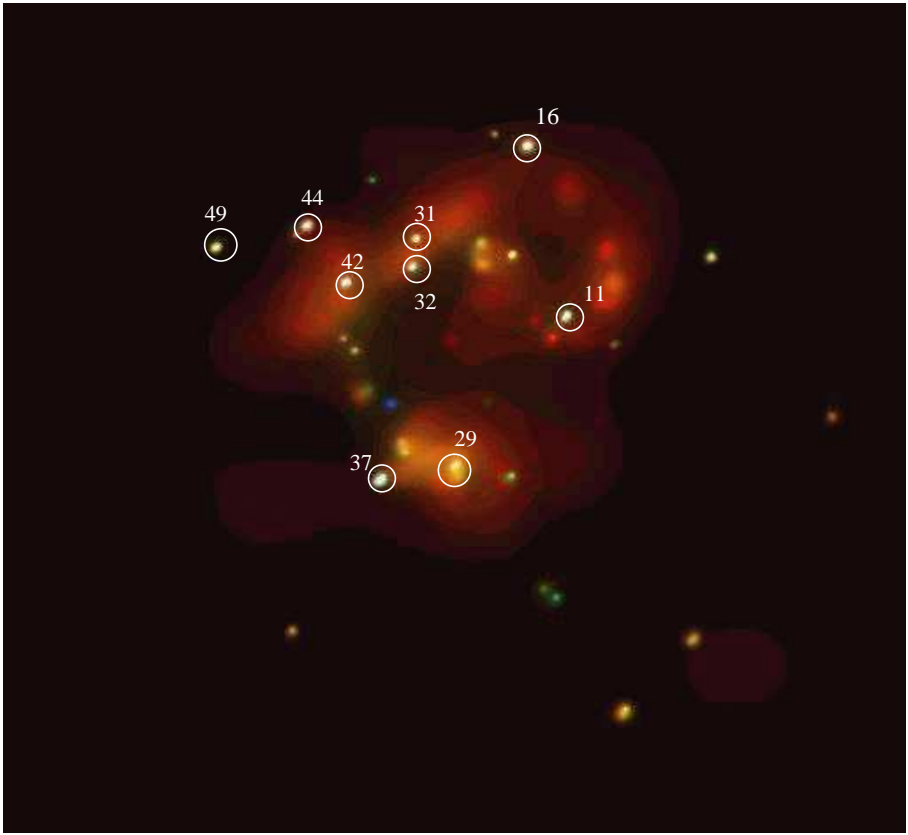}
\caption{The Dec 1999 image of The Antennae (Fabbiano et al.\ 2001), with 
the ULXs discussed in this paper identified by circles and source
numbers from Zezas et al.\ (2002a). Soft emission is red, hard emission is blue.
The ULXs appear white because they emit in the entire spectral band.}
\end{figure}

\begin{figure}
\begin{tabular}{cc}
\rotatebox{0}{\includegraphics[width=7.0cm]{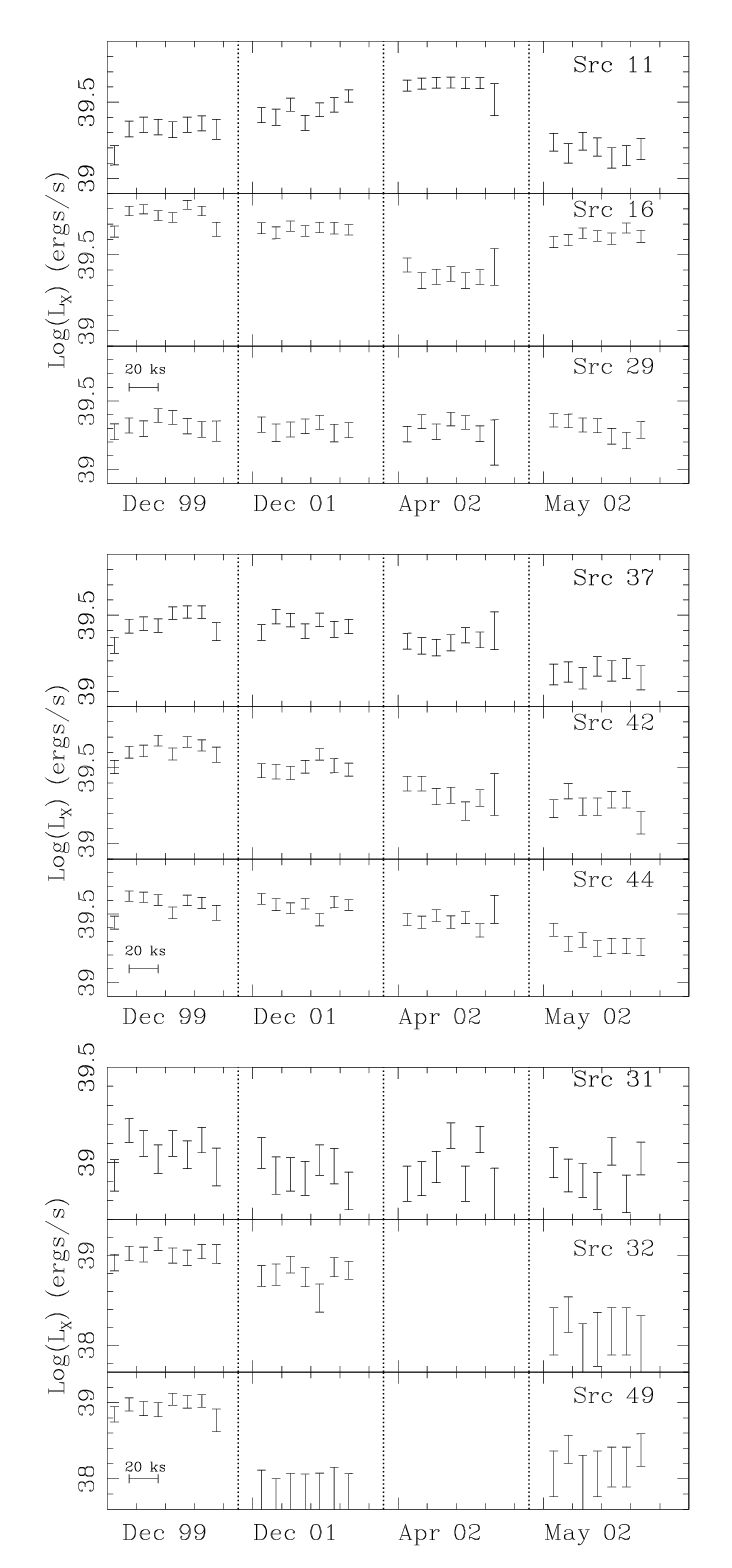}}
\end{tabular}
\caption{Light curves of the ULXs, identified by the source numbers in Zezas
et al.\ (2002a). 
The horizontal axes are labeled with the date of the observation (see Table~1).
A 20~ks time bar is also shown in each panel.}
\end{figure}

\begin{figure}
\begin{tabular}{cc}
\rotatebox{0}{\includegraphics[width=9.0cm]{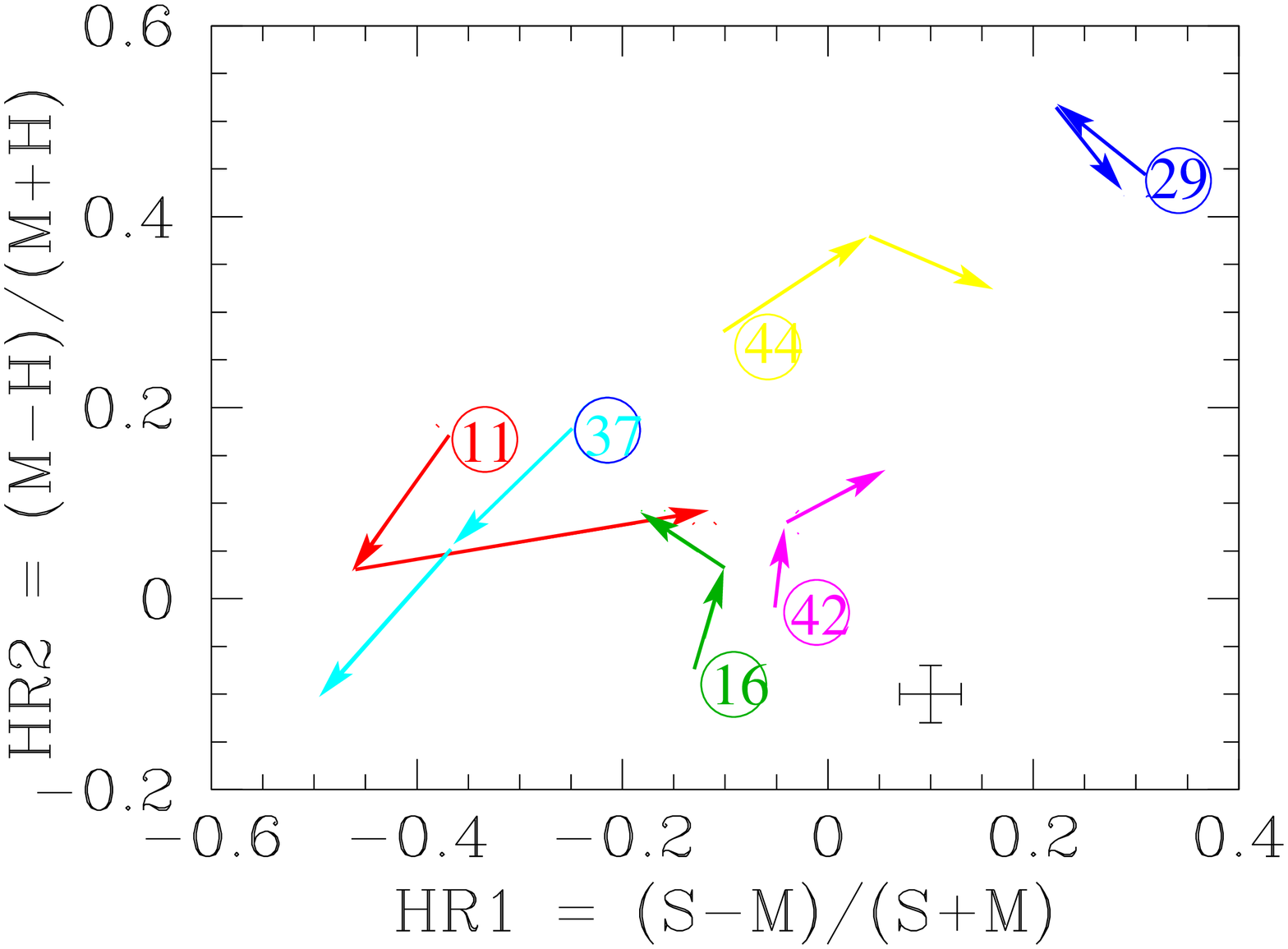}} & \\
\rotatebox{0}{\includegraphics[width=9.0cm]{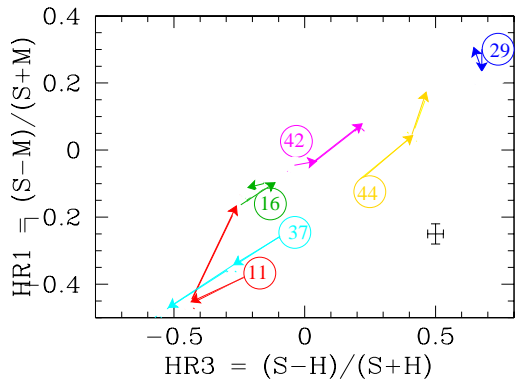}} & \\
\end{tabular}
\caption{X-ray color--color tracks of ULXs. The source number identifies the first point in time.
The arrow shows the direction of time.}
\end{figure}


\begin{thebibliography}{}


\bibitem[Angelini, Loewenstein,  Mushotzky(2001)]{2001ApJ...557L..35A} 
Angelini, L., Loewenstein, M., \& Mushotzky, R.~F. 2001, \apjl, 557, L35 

\bibitem[Colbert \& Ptak]{} Colbert, E. J. M. \& Ptak, A. F. 2002, \apjs, in press
(astro-ph/0204002)

\bibitem[Esin et al 1998]{}Esin, A., Narayan, R., Cui, W., Grove, J. E., \& Zhang, S. N. 1998,
\apj, 505, 854

\bibitem[Fabbiano 1989]{f89} Fabbiano, G. 1989, Ann. Rev. Ast. Ap., 27, 87


\bibitem[Fabbiano et al 2001]{}Fabbiano, G., Zezas, A., \& Murray, S.
2001, \apj, 554, 1035 

\bibitem[Fabian \& Terlevich 1996]{}Fabian, A. \& Terlevich, R. 1996, MNRAS, 280, 5

\bibitem[Frank et al 2002]{}Frank, J., King A.R., Raine, D.J., 2002,
Accretion Power in Astrophysics (3rd ed; Cambridge: Cambridge University Press)

\bibitem[Greiner at al 2002]{}Greiner, J., Cuby, J. G.,
 McCaughrean, M. J. 2001, Nature, 414, 522


\bibitem[Janiuk, Zycki \& Czerny 2000]{}Janiuk, A., Zycky, P. T. \& 
Czerny, B. 2000, MNRAS, 314, 364

\bibitem[Kilgard et al.(2002)]{} Kilgard, R.,
Kaaret, P., Krauss, M.,  Prestwich,  A.~H., Raley, M. \& Zezas, A.,
2002,  \apj, 573, 138

\bibitem[King 2002]{}King, A. R. 2002, MNRAS, 335, L13

\bibitem[King et al.(2001)]{2001ApJ...552L.109K} King, A.~R., Davies, 
M.~B., Ward, M.~J., Fabbiano, G., \& Elvis, M.\ 2001, \apjl, 552, L109

\bibitem[Kong et al 2002]{}Kong, A. K. H., Garcia, M. R.,
 Primini, F. A., Murray, S. S.,
 Di Stefano, R.,
 McClintock, J. E. 2002, \apj, 557, 738
 
\bibitem[Kubota et al.(2001)]{2001ApJ...547L.119K} Kubota, A., Mizuno, T., 
Makishima, K., Fukazawa, Y., Kotoku, J., Ohnishi, T., \& Tashiro, M.\ 2001, 
\apjl, 547, L119 

\bibitem[La Parola et al.(2001)]{2001ApJ...556...47L} La Parola, V., Peres, 
G., Fabbiano, G., Kim, D.~W., \& Bocchino, F.\ 2001, \apj, 556, 47 

\bibitem[La Parola et al.(2002)]{}La Parola, V., Damiani, F., Fabbiano, G.
and Peres, G. 2002, \apj, in press (astro-ph/0210174)

\bibitem[Makishima \etal\ 2000]{}Makishima, K. \etal\ 2000, \apj, 535,
632

\bibitem[Miller et al 2002]{}Miller, J. M., Wijnands, R., Rodriguez-Pascual, P. M.,
Ferrando, P., Gaensler, B. M., Goldwurm, A., Lewin, W. H. G., and Pooley, D
2002, \apj, 566, 358

\bibitem[Misra \& Sriram 2002]{}Misra, R. \& Sriram, K. 2002, preprint
(astro-ph/0210457)

\bibitem[Neff \& Ulvestad 2000]{} Neff, S. G. \& Ulvestad, J. S. 
2000, \aj, 120, 670

\bibitem[Pringle 1997]{}Pringle, J. E. 1997, MNRAS, 292, 136

\bibitem[Shapiro, Lightman \& Eardley 1976]{}Shapiro, S., Lightman, A., 
\& Eardley, D. 1976, \apj, 204, 187

\bibitem[Smith, Heindl, \& Swank 2002]{} Smith, D. M., Heindl, W. A.,
\& Swank, J. H. 2002, \apj, 569, 362

\bibitem[Stark et al.(1992)]{1992ApJS...79...77S} Stark, A.~A., Gammie, 
C.~F., Wilson, R.~W., Bally, J., Linke, R.~A., Heiles, C., \& Hurwitz, M.\ 
1992, \apjs, 79, 77 

\bibitem[Weisskopf \etal\ 2000]{}Weisskopf, M., Tananbaum, H., Van Speybroeck, L. \& O'Dell, S.
2000, Proc. SPIE 4012 (astro-ph 0004127)

\bibitem[Whitmore et al.(1999)]{1999AJ....118.1551W} Whitmore, B.~C., 
Zhang, Q., Leitherer, C., Fall, S.~M., Schweizer, F. \& Miller, B.~W.\ 
1999, \aj, 118, 1551 

\bibitem[Zezas \& Fabbiano 2002]{} Zezas, A. \& Fabbiano 2002, \apj, 577, 726

\bibitem[Zezas \etal\ 2001]{} Zezas, A., Fabbiano, G. Rots, A. H.,\&
Murray, S., 2002a, \apjs, 142, 239

\bibitem[Zezas \etal\ 2001]{} Zezas, A., Fabbiano, G. Rots, A. H.,\&
Murray, S., 2002b, \apj, 577, 710


\end{thebibliography}
\end{document}